\documentstyle[12pt,twoside]{article}
\begin{document}

\title{\large\bf Generalised scalar-tensor theory and the cosmic acceleration}

\author{Narayan Banerjee{\footnote {E-mail: narayan@juphys.ernet.in ;}}~ and 
Koyel Ganguly{\footnote{E-mail: koyel\_g\_m@yahoo.co.in}} \\ Relativity and 
Cosmology Research Centre,\\ Department of Physics, Jadavpur University,\\
Kolkata - 700032,\\India.}

\date{}

\maketitle
\vspace{0.5cm}
{\em PACS Nos. : 04.20.-q, 98.80.Jk}
\vspace{0.5cm}

\pagestyle{myheadings}
\newcommand{\be}{\begin{equation}}
\newcommand{\ee}{\end{equation}}
\newcommand{\bea}{\begin{eqnarray}}
\newcommand{\eea}{\end{eqnarray}}

\begin{abstract}
In this paper it has been shown that a simple functional form of $\omega(\phi)$ in a generalised scalar tensor theory 
can drive the present cosmic acceleration without any quintessence field or the cosmological constant $\Lambda.$ 
Furthermore, it ensures a smooth transition from a decelerated to an accelerated phase of expansion in the matter 
dominated regime. 
\end{abstract}

\section{Introduction}
Although high precision observational data and their interpretations point 
towards an accelerated expansion of the universe with more and more 
certainty\cite{WMAP}, the search for the `dark energy' sector, which 
drives this acceleration, has not seen a preferred direction as yet. The 
good old cosmological constant $\Lambda$, the most proclaimed candidate as 
the source for this repulsive gravitational effect, does fit the 
observational data reasonably well, but it has its own problems\cite{VS}. 
 Naturally a large number of other alternatives have already appeared in 
the literature with their own virtues and shortcomings\cite{Review}. 
\par Albeit their problems with local astronomical experiments, 
non-minimally coupled scalar field theories, particularly in the framework 
of Brans-Dicke (BD) theory, have proved to be useful in negotiating this 
counter-intuitive acceleration. It has been shown that BD theory along with a quintessence scalar 
field can indeed generate an accelerated expansion of the universe\cite{NB-DP}. A variation of BD theory, for example an
addition of a potential $V$ which is a function of BD scalar field itself, can drive 
this desired accelerated expansion\cite{BM}. Most of these models suffer 
from two important drawbacks. One is that in these models the matter 
dominated universe has an ever accelerating expansion contrary to the 
recent observations\cite{RS} as well as theoretical requirements\cite{PT}.
In a recent work, however, it has been shown that along with a 
quintessence scalar field which interacts with the BD field, it is 
possible to have a scenario in which the quintessence field oscillates at 
an early epoch but grows later to drive the accelerated expansion for a 
fairly arbitrary set of quintessence potentials\cite{MP}.
\par The second problem is of an entirely different nature. The 
dimensionless parameter $\omega$ in Brans-Dicke theory plays a crucial 
role in the prediction of observational results. Although the 
popular belief that BD theory goes over to GR in the infinite $\omega$ 
limit suffered a jolt\cite{SS}, but in the weak field regime BD results 
get closer to GR results for higher values of $\omega$. The local 
astronomical experiments 
are quite well explained by GR and demands a pretty high (a few hundreds) 
value of $\omega$\cite{WILL} if the predictions made in BD theory have to 
be within the observational uncertainty. On the other hand in most of the 
models in the Brans-Dicke 
framework, the accelerated expansion of the universe requires a very low 
value of $\omega$, typically of the order of unity. However, a recent work 
shows that if the BD scalar field interacts with the dark matter, a 
generalised BD theory can perhaps serve the purpose of driving an 
acceleration even with a high value of $\omega$\cite{SD}.
\par In these investigations, either Brans-Dicke theory is modified to 
suit the present requirement or a quintessence scalar field is used to 
generate sufficient acceleration. In ref.\cite{SD} and in a recent work 
by Barrow and Clifton\cite{BC}, no additional potential were added, but an 
interaction between the BD scalar field and the dark matter were used to do 
the needful.
\par It was also shown that a Brans-Dicke scalar field alone can drive an 
accelerated expansion in the matter dominated epoch, without any 
quintessence matter or any interaction between the BD field and the dark 
matter\cite{DP}. The problem once again was that it required a very low 
value of $\omega$, of the order of unity, and there was no transition from 
a decelerated to an accelerated scenario.  
\par In the present work, we intend to show that a generalisation of 
Brans-Dicke theory by Bergman and Wagoner\cite{BW} and in a more useful 
form by Nordtvedt\cite{Nd} can in fact solve at least the first problem. 
In this generalisation, the parameter $\omega$ is taken to be a 
function of the BD scalar field instead of its being a constant. Different 
functional forms of $\omega$ could originate from various physical 
motivations. It is indeed an appealing feature of any model if the 
accelerated expansion can be generated without the requirement of an 
additional quintessence field. Naturally it would be interesting to check 
if some form of $\omega(\phi)$ can give rise to a decelerated expansion to 
start with and helps entering into an accelerated expansion phase later, 
but all in the matter dominated regime. In what follows we shall show that 
indeed a simple choice of $\omega$ as a function the Brans-Dicke scalar field 
does the trick.
\par In the next section the model with a variable $\omega$ is 
presented. Section $3$ deals with some specific examples where the 
deceleration parameter has a smooth transition from a positive to a 
negative value and section $4$ presents some discussions on the results 
obtained and the possibilities for some future work. 
  \section{Field Equations}
For a spatially flat Robertson Walker spacetime, the field equations in 
the generalised Brans-Dicke theory are,
\bea
3\frac{\dot{a}^2}{a^2}= 
\frac{\rho}{\phi}+\frac{\omega(\phi)}{2}\frac{\dot{\phi}^2}{\phi^2}-3\frac{\dot{a}}{a}\frac{\dot{\phi}}{\phi},\\
2\frac{\ddot{a}}{a}+\frac{\dot{a}^2}{a^2}= 
-\frac{\omega(\phi)}{2}\frac{\dot{\phi}^2}{{\phi}^2} - 2\frac{\dot{a}}{a}
\frac{\dot{\phi}}{\phi}-\frac{\ddot{\phi}}{\phi},
\eea
where $a$ is the scale factor of the universe, $\rho$ is the density of the 
matter distribution, $\phi$ is the Brans-Dicke scalar field and $\omega$ the 
dimensionless parameter, now a function of $\phi$ rather than being a 
constant. The units are chosen where $8{\pi}G = 1.$ The thermodynamic pressure of the cosmic fluid is taken to be 
zero consistent with the present dust universe. In what follows, we shall 
assume the conservation equation for matter leading to the relation
\be
\rho = \frac{\rho_0}{a^3},\\
\ee
where $\rho_0$ is a constant. Hence the wave equation for the scalar field 
$\phi$,\\
\be
\ddot{\phi} + 3\frac{\dot{a}}{a}\dot{\phi} = 
\frac{1}{2\omega+3}(T-\dot{\phi}^2\frac{d\omega}{d\phi}),\\
\ee
is not an independent equation, and follows from the Bianchi identities. 
Here $T$ is the trace of the energy momentum tensor for matter. So 
we have two equations $(1)$ and $(2)$ for three unknowns $a$, $\phi$ and 
$\omega$. We choose the relation,\\
\be
\phi = \phi_0 a^n,\\
\ee
so that the equation system is now closed. Here $\phi_0$ and $n$ are 
constants. There is however no apriori reason for this choice only except 
that the equation system becomes tractable and consistent. With this 
choice and equation $(3)$, the system of equations can be readily 
integrated for a constant $\omega$ corresponding to Brans-Dicke theory. The 
scale factor $a(t)$ comes out as a simple power function of the cosmic 
time $t$ and hence rules out the possibility of any transition of the 
deceleration parameter $q = 
-\frac{\ddot{a}a}{\dot{a}^2}$ from a positive to a negative value in the 
matter dominated epoch. The expansion can either be ever accelerating or ever 
decelerating depending on the choice of values of the constants. This is in 
perfect agreement with the work already in the literature\cite{SD} and \cite{DP}. In Nordtvedt's 
generalised theory with a varying $\omega$, the situation can be 
dramatically different. Using relation $(5)$ in a combination of equations 
$(1)$ and $(2)$, one can write\\
\be
\frac{\ddot{a}}{a}+(n+2)\frac{\dot{a}^2}{a^2}=\frac{\rho_0}{\phi_0}a^{-(n+3)},\\
\ee
which has a first integral of the form\\
\be
\dot{a}^2=\frac{2\rho_0}{\phi_0(n+2)(n+3)}a^{-(n+1)}+{A}a^{-2(n+2)},\\   
\ee
$A$ being a constant of integration.
From equations $(6)$ and $(7)$, the deceleration parameter $q$ can be written as\\
\be
q = -\frac{\ddot{a}a}{a^2} = \frac{A\phi_0{(n+2)}^2(n+3)+\rho_0(n+1)a^{(n+3)}}{A\phi_0(n+2)(n+3)+2\rho_0a^{(n+3)}}.\\
\ee
Amongst the constants $\rho_0$ is positive definite and so is $\phi_O$ as the Newtonian constant of gravitation and 
hence $\phi$ should always be positive. The constants $A$ and $n$ can have negative values as well and can help finding 
a $q$ which gets into a negative value in a recent past, if certain conditions are satisfied.

\section{A particular model}
\par In what follows, we shall assume that $A$ is positve but $-3<n<-1$. This indeed gives a possibility that $q$ starts 
positive for small $a$ but eventually attains a negative value when\\
$$|n+1| \rho_0 a^{(n+3)} > A \phi_0 {(n+2)}^2 (n+3).$$\\
In order to get a flair for the numbers, we choose\\
$$n = -\frac{3}{2},$$\\
so that $q = \frac{1}{2}$ for a negligible value of $a,$ i.e, we assume at the outset that the beginning of the matter 
dominated $(p = 0)$ era behaves the same as that in spatially flat FRW model in general relativity. The relation 
between the constants will now determine the time at which $q$ crosses the zero value. If we assume that $q = 0$ at $z 
 = 1.5$ we get the relation\\
\be
\frac{3A\phi_0}{4} = \rho_0{(2/5)}^{3/2},\\
\ee
where the redshift $z$ is given by $$1+z=\frac{a_0}{a},$$ the subscript 0 
indicating the present value.
The deceleration parameter then has a simple form\\
\be
q = 
\frac{1}{2}[\frac{1-{(\frac{5a}{2})}^{3/2}}{1+2{(\frac{5a}{2})}^{3/2}}].\\
\ee
We have scaled $a$ such that its present value $a_0 = 1.$ Thus, for $n = 
-\frac{3}{2},$ the deceleration parameter $q$ is close to $0.5$ for a very 
small value of $a,$ becomes zero at $a=\frac{2}{5},$ i.e, $z = 1.5$ and 
has a negative value of $q \approx - 0.16$ at the present epoch. Clearly, the 
time of transition of the signature of $q$ is sensitive to the choice of the 
constants, and hence a fine tuning will enable us to get the correct epoch 
where $q$ crosses the zero value. \\
\par Using equations $(3),(5)$ and $(7)$ in equation $(1)$ with the choice 
$n = -\frac{3}{2},$ the functional dependence of the choice of 
$\omega(\phi)$ can be written as\\
\be
\omega(\phi) = -\frac{4}{3}[1+\frac{1}{2(2+\phi)}].\\
\ee
Clearly $\omega$ has a negative value, which indicates where does the 
negative contribution to the effective pressure come from. For other 
choices of the constant $n,$ the functional form of $\omega(\phi)$ will 
be different. Furthermore, the form of $\omega(\phi)$ also fine tunes the 
value of $z$ at which $q$ crosses the zero value in favour of a negative 
one. For instance, if $n = -\frac{3}{2}$ and $\omega(\phi)$ is given as\\
\be
\omega(\phi) = -\frac{4}{3}[1+\frac{\alpha}{2(\phi+2\alpha)}],\\
\ee
where $$\alpha = {(2)}^{3/2},$$\\
the signature flip in $q$ takes place at $z = 1.0.$ With the same form of $\omega(\phi)$ with $\alpha = {(3/2)}^{3/2},$ 
the flip takes place at $z = 0.5.$

\par  In view of the high degree of non-linearity of Einstein's equations, it is
 now important to check whether the reconstruction of the form of 
 $\omega = \omega (\phi)$ gives rise to the same form of $q$. In this work, 
 the detailed stability analysis, presumably the subject matter for another 
 full-fledged paper, is not carried out. But it can be said that the forms 
 of $\omega(\phi)$ given are both necessary and sufficient for the 
 corresponding behaviour of $q$. For example, a particular evolution for $q$ 
 given by equation (10) yields the form of $\omega(\phi)$ given by equation 
 (11) showing the necessity of the latter to arrive at the form of $q$ given 
 by equation (10). However, if one now takes up (11) as the input and use 
 equations (2) and (5), the same behaviour of $q$ is obtained. This shows 
 the sufficiency of equation (11). So long as the results of the stability 
 analysis is not known, this necessary and sufficient nature of equation 
 (11) shows that the solution does worth attention.

\section{Discussions:}
The present work clearly shows that a generalised scalar tensor theory 
where the BD parameter $\omega$ is a function of the scalar field $\phi,$ 
can drive an accelerated expansion for the present universe without having 
to resort to an additional quintessence field. Unlike most of the 
Brans-Dicke models, a varying $\omega$ even allows for a signature flip in 
$q$ in the matter dominated epoch. This indeed requires a fine tuning of 
the parameters, but the merit of the model is that this transition can be 
shown analytically.
\par The value of $\omega,$ which effects this smooth transition, is not 
specified, only the functional form of $\omega$ can be determined. But 
this gives an advantage. For local astronomical experiments, $\omega$ can 
have a high value due to the local inhomogeneity as $\phi$ would be 
function of the space coordinates. But at a cosmological scale, the value 
of $\omega,$ averaged over the spatial volume of the universe, could be 
small and hence one can avoid the nagging problem of the discrepancy of 
the values of $\omega$ for a cosmological requirement and the local 
experiments.
\par The present work, however, has its own problems. Although the value 
of $\omega$ required is not specified, equation $(12)$ indicates that it
 has a low negative value. This contradicts the local astronomical requirement
 of a high value of $\omega$ as mentioned earlier. Furthermore, a negative 
$\omega,$ particularly $\omega < -\frac{3}{2}$ leads to a negative
 contribution to the kinetic part of the energy leading to quantum instabilities\cite{FP}. However this problem is shared by most of the phantom models with 
a negative Hamiltonian. The form of $\omega(\phi)$ is chosen phenomenologically
 rather than inspired by any underlying physics. For that matter the
 quintessence potentials are all chosen like that so the present model is no 
 worse than any of the quintessence models on the count of a sound theoretical 
 basis. The advantage here is that the scalar field itself is already
 there in the purview of the theory and is not put in by hand.
\par The particular model presented in section $2$ assumes $\phi$ as a 
power function of the scale factor as $\phi = \phi_0{a}^n,$ which yields\\
$$\frac{\dot{\phi}}{\phi} = nH.$$\\
As in this theory, $\phi$ is the inverse of the effective Newtonian 
constant $G,$ thus one has\\
$$\frac{\dot{G}}{G} = -nH.$$\\
For this particular model to work efficiently one requires that $n$ should 
be of the order of unity, so the fractional rate of variation of $G$ is of 
the same order of magnitude as $H.$ Observational limits indicate that the 
rate should be smaller\cite{VG}. Definitely one would have been more 
comfortable with values of $n$ not greater than ${10}^{-1},$ but this is 
only a primitive model, and the high degree of nonlinearity keeps the 
possibility of getting the required features of the model with other 
choices of $\omega(\phi)$ wide open. Surely investigations along this 
line, i.e, to find a form of $\omega(\phi)$ which preserves the features 
of this model and gives a better value of $\frac{\dot{G}}{G},$ is 
warranted.
\par The other problem of the model is quite generic for all the dark energy
 candidates, namely that of the fine tuning of parameter. One exception of 
this is of course the tracking solutions where the potential grows to drive
 acceleration in the later stages from a wide range of initial conditions 
 \cite{stein}.

\vspace{0.5cm}

{\large\bf Acknowledgement:}\\

\par The authors thank the BRNS (DAE) for financial support. We also thank
 the anonymous referee for some useful suggestions. \\

\end{document}